\documentclass[12pt]{article}
\input psfig.sty

\makeatletter
\renewcommand\section{\@startsection {section}{1}{\z@}%
                                   {-3.5ex \@plus -1ex \@minus -.2ex}%
                                   {2.3ex \@plus.2ex}%
                                   {\normalfont\large\bfseries}}
\renewcommand\subsection{\@startsection{subsection}{2}{\z@}%
                                     {-3.25ex\@plus -1ex \@minus -.2ex}%
                                     {1.5ex \@plus .2ex}%
                                     {\normalfont\normalsize\bfseries}}
\renewcommand\subsubsection{\@startsection{subsubsection}{3}{\z@}%
                                     {-3.25ex\@plus -1ex \@minus -.2ex}%
                                     {1.5ex \@plus .2ex}%
                                     {\normalfont\normalsize\bfseries}}
\long\def\@makecaption#1#2{%
  \vskip\abovecaptionskip
  \normalbaselineskip=12pt\baselineskip=12pt
  \sbox\@tempboxa{\small #1: #2}%
  \ifdim \wd\@tempboxa >\hsize
    {\small #1: #2}\par
  \else
    \global \@minipagefalse
    \hb@xt@\hsize{\hfil\box\@tempboxa\hfil}%
  \fi
  \vskip\belowcaptionskip}
\makeatother

\def\mpalla{\vbox{\normalbaselineskip=0pt\baselineskip=0pt%
\hbox{\null\kern0.25em$\scriptstyle\circ$}\hbox{$m$}}}

\def\cG{{\cal G}}
\def\cS{{\cal S}}

\def\cI{{\cal I}}
\def\cT{{\cal T}}

\newcommand\0{\hphantom{0}}
\newcommand\half{\textstyle{1\over2}}

\begin{document}
\date{{\normalsize May 8, 2000}}
\title{
\null\vskip -60 pt
\begin{flushright}
\normalsize\rm IFUP-TH/2000-13
\end{flushright}
\vskip 10 pt
\Large\bf Linked-Cluster Expansion of the Ising Model}
\author{\Large Massimo Campostrini\\
\normalsize {\it INFN, Sezione di Pisa, 
and Dipartimento di Fisica dell'Universit\`a di Pisa}\\
\normalsize {\tt Massimo.Campostrini@df.unipi.it}}
\maketitle

\begin{abstract}
The linked-cluster expansion technique for the high-temperature
expansion of spin model is reviewed.  A new algorithm for the
computation of three-point and higher Green's functions is presented.
Series are computed for all components of two-point Green's functions
for a generalized $3D$ Ising model, to 25th order on the bcc lattice
and to 23rd order on the sc lattice.  Series for zero-momentum four-,
six-, and eight-point functions are computed to 21st, 19th, and 17th
order respectively on the bcc lattice.
\end{abstract}

\section{Introduction}
\label{Introduction}

The high-temperature (strong-coupling) series expansion is one of the
most successful tools for the study of physical systems near a
critical point.

High-temperature series are analytic; the radius of convergence is
usually quite large, often reaching the boundary of the
high-temperature phase.  This property allows the application of
powerful techniques of resummation and analytical continuation
\cite{Guttmann}, which can yield very precise and reliable results,
provided that long series are available.  It is therefore worthwhile
to push the computation of high-temperature series as far as our
algorithms and computers allow.

The most successful technique for the computation of high-temperature
series of $3D$ spin models is the linked-cluster expansion (LCE),
which is well suited for the fully computerized approach required to
reach very high orders of the expansion.

Several detailed discussion of the LCE appeared in the literature.
Wortis' review \cite{Wortis} covers most of the basic topics, and
provides many graphical rules fit for algorithmic implementation.
Nickel performed a remarkable computation for the $3D$ Ising model on
the bcc lattice \cite{Nickel-Cargese}, which had not been surpassed
until the present work; in a more detailed paper, Nickel and Rehr also
present several clever algorithms which we found very useful
\cite{Nickel-Rehr}.  L\"uscher and Weisz, describing their
application of the LCE to lattice field theory, also provide several
important implementation hints \cite{Luscher-Weisz}.

Unfortunately, the notation found in the literature is by no means
uniform.  Therefore we will review the relevant aspects of the LCE,
which can be found in Ref.\ \cite{Wortis}, not only to make our paper
more self-contained, but also to explain notations carefully and to
remark the correspondence with Refs.\ 
\cite{Wortis,Nickel-Rehr,Luscher-Weisz}.  We will follow the notations
of Ref.\ \cite{Nickel-Rehr} whenever possible.

The paper is organized as follows:

Sect.\ \ref{Graphology} introduces the relevant graph theory
concepts and definitions.

Sect.\ \ref{Model} presents the generalized Ising model we focus on.

Sects.\ \ref{Unrenorm}, \ref{1-renorm}, and \ref{2-renorm} review the
LCE, with special focus on the two-point Green's functions.

Sect.\ \ref{q-point} describes our algorithm for the computation of
three-point and higher Green's functions.

Sect.\ \ref{Programming} is devoted to programming details.

Sect.\ \ref{Results} displays a (small) selection of the series
generated.

Forthcoming papers will be devoted to the analysis of the series,
using the techniques presented in Ref.\ \cite{CPRV-Ising-es}, and to
the generation and analysis of the series for $XY$ systems.

We will not give proofs of our formulae.  The only nontrivial step in
the proofs of Sects.\ \ref{1-renorm}--\ref{q-point} is to show that
the symmetry factors compensate exactly the different number of
contributions that may appear on the two sides of the equations; it is
typically a straightforward, if tedious, exercise in combinatorics.
The proofs of Sects.\ \ref{Unrenorm}--\ref{2-renorm} are given or
sketched in Ref.\ \cite{Wortis}.  The proofs of Sect.\ \ref{q-point}
are especially easy, since the symmetry factor of a 1-irreducible tree
graph is always 1.

\section{Graphology}
\label{Graphology}

In this section we introduce a number of graph theory concepts
relevant for the LCE.  We refer the reader to Ref.\
\cite{Essam-Fisher} for a comprehensive introduction to the subject.

A {\em graph\/} is a set of {\em vertices\/} and {\em edges\/} (also
named {\em links\/} or {\em bonds\/} in the literature).  Each edge
$l$ is {\em incident\/} with two distinct vertices, its extrema (we do
not allow the extrema to coincide); the set of extrema will be denoted
by $\partial l$; we write $\partial l = \{i(l),f(l)\}$; the choice of
an ``initial'' and a ``final'' vertex is arbitrary.  Two vertices are
{\em adjacent\/} if they are the extrema of the same edge.  We will
denote the number of vertices and edges of a graph by $v$ and $e$
respectively.  We also consider {\em arcs\/} or {\em oriented edges},
incident out of the initial vertex $i(l)$ into the final vertex
$f(l)$.

The {\em valence} $n(i)$ of a vertex $i$ is the number of edges
incident with $i$.

An $r$-{\em rooted\/} graph is a graph with $v$ vertices, $v \ge r$:
$r$ {\em roots\/} or {\em external vertices} and $v-r$ {\em internal
vertices}.  We will assign the indices $1$, ..., $r$ to the roots and
the indices $r+1$, ..., $v$ to the internal vertices.  In the
drawings, roots will be denoted by open dots and internal vertices by
filled dots.

Two $r$-rooted graphs are {\em isomorphic\/} if there exists a
one-to-one correspondence $\pi$ of their internal vertices and edges
such that the incidence relations are preserved, i.e.,
$\partial(\pi(l)) = \{\pi(i(l)),\pi(f(l))\}$ ($\pi(i)=i$ for the
roots). From now on, we will identify isomorphic graph, and silently
assume that all sets of graphs we define contain only non-isomorphic
graphs.

The {\em symmetry factor\/} $S(\cG)$ of an $r$-rooted graph $\cG$ is
the number of isomorphisms of $\cG$ into itself, i.e.\ the number of
of permutations of internal vertices and edges preserving the
incidence relations.

We will also consider {\em $p$-ordered $r$-rooted graphs}, $p \le r$,
i.e.\ the classes of rooted graphs isomorphic up to permutations of
$r-p$ roots.  We will assign indices $1$, ..., $p$ to the fixed roots
and indices $p+1$, ..., $r$ to the roots which can be permuted.  The
symmetry factor $S(\cG)$ is defined as the number of isomorphisms of
$\cG$ into itself, including permutation of the roots $p+1$, ..., $r$.
The symmetry factor divided by $(r-p)!$ is called the {\em modified
symmetry factor\/} $S_E(\cG)$ (cfr.\ Ref.\ \cite{Luscher-Weisz}); it
need not be an integer.  The $r$-rooted graphs defined above are 
{\em ordered\/} ($r$-ordered); unless otherwise specified, we will
assume that graphs are ordered.  0-ordered graphs are {\em unordered}.

We will also discuss ($p$-ordered) $r$-rooted graphs whose edges
and/or vertices are assigned a label; let us consider e.g.\ the case
of an edge label $a(l)$ and vertex label $b(i)$.  Two labelled graphs
$(\cG,a,b)$ and $(\cG',a',b')$ are equivalent (isomorphic) if there
exists an isomorphism of $\cG$ into $\cG'$ such that $a$ is mapped
into $a'$ and $b$ is mapped into $b'$.  The symmetry factor
$S(\cG,a,b)$ is the number of isomorphisms of $(\cG,a,b)$ into itself.

A pair of vertices $i$ and $j$ is {\em connected\/} if there exists a
sequence of vertices $k_1$, ..., $k_n$, with $k_1=i$ and $k_n=j$, and
a sequence of edges $l_1$, ..., $l_{n-1}$ such that $\partial l_a =
\{k_a,k_{a+1}\}$, $a=1$, ..., $n-1$.  A graph is {\em connected\/} if
every pair of its vertices is connected.  In the following, unless
otherwise noted, we will assume that every graph is connected.

A sequence of distinct vertices $k_1$, ..., $k_n$ and distinct edges
$l_1$, ..., $l_n$ is called a {\em loop\/} of length $n$ if $\partial
l_a = \{k_a,k_{a+1}\}$, $a=1$, ..., $n-1$, and $\partial l_n =
\{k_n,k_{1}\}$.  The number of independent loops (also known as
{\em cyclomatic number\/}) of a connected graph is $e - v + 1$.

A connected graph is called a {\em tree graph\/} if it contains no
loop.  A tree graph has $v-1$ edges.

\section{The model}
\label{Model}

We wish to compute the high-temperature (HT) expansion of the
$q$-point functions of a generalized Ising model on a $D$-dimensional
Bravais lattice $\Lambda$; notice that Bravais lattices enjoy
inversion symmetry at each lattice site.  The model is defined by the
generating functional
\begin{equation}
\exp(W[h]) = {1 \over Z} \prod_i \left[\int d\phi_i\,
f(\phi_i) \exp(h_i \phi_i)\right]
\exp\Biggl(K \sum_{\langle ij \rangle} \phi_i \phi_j\Biggr),
\label{eq:S}
\end{equation}
where $\phi_i$ is a scalar field, $f$ is an even non-negative function
or distribution decreasing faster than $\exp(-\phi^2)$ as
$\phi\to\infty$, normalized by the condition
\[
\int d\phi\, f(\phi) = 1,
\]
$K=\beta J$, the sum runs over all pairs of nearest neighbours, and
the normalization $Z$ is fixed by requiring $W[0] = 0$.

The connected $q$-point function (denoted by $\cal M$ in
Ref.\ \cite{Wortis}) at zero magnetic field is defined by
\begin{equation}
G_q(x_{i_1},...,x_{i_q}) \equiv 
\left.\langle\phi_{i_1} ... \phi_{i_q}\rangle^{\rm con}\right|_{h=0} =
\left. \partial^q W[h] \over \partial h_{i_1} \, ... \, 
       \partial h_{i_q} \right|_{h=0},
\end{equation}
where $x_i \equiv (x_1(i),...,x_D(i))$ is the coordinate vector of the
lattice site $i$.  $G_q$ is invariant under the lattice symmetry
group, including (discrete) translations, and under permutation of its
arguments; it is customary to write $G_2(x_{i_1},x_{i_2})$ in the form
$G_2(x_{i_1}-x_{i_2})$.  We will apply the LCE to the computation of
$G_q$.

\section{Unrenormalized expansion}
\label{Unrenorm}

Let us parametrize the potential $f$ in terms of the bare vertices
$\mu_0(2n)$, defined by the generating function
\begin{equation}
\exp\left[\sum_n {\mu_0(2n)\over(2n)!}\,h^{2n}\right] =
\int d\phi\, f(\phi) \exp(h \phi).
\end{equation}
These quantities are named bare semi-invariants
and denoted by ${M_n}^0$ in Ref.\ \cite{Wortis}; they are named
cumulant moments and denoted by $\mu_{2n}$ in Ref.\ 
\cite{Nickel-Rehr}; $\mu_0(2n) = (2n-1)!!\,\mpalla^{\rm con}_{2n}$ in
the notations of Ref.\ \cite{Luscher-Weisz}.  
Without loss of generality, we can rescale $\phi$ and $K$ to fix 
$\mu_0(2)=1$.

For a generic $r$-rooted graph $\cG$, we define the bare external vertex
factor
\begin{equation}
V_0^{(e)}(n_1,...,n_r;\cG) = \prod_{i=1}^r \mu_0(n(i)+n_i),
\label{V0e}
\end{equation}
the bare internal vertex factor
\begin{equation}
V_0^{(i)}(\cG) = \prod_{i=r+1}^v \mu_0(n(i)),
\label{V0i}
\end{equation}
and the bare edge factor
\begin{equation}
L_0(x_1,...,x_v; \cG) = \prod_{l=1}^e [K \theta(x_{i(l)}-x_{f(l)})],
\label{L-0}
\end{equation}
where $\theta(x) = \delta(\Vert x \Vert -1)$ and $\Vert x \Vert$ is
the lattice distance between 0 and $x$.  

It is convenient to focus on the contributions of $r$-rooted graphs to
$q$-point functions.  To this purpose we introduce the auxiliary
$r$-point functions $X$, whose unrenormalized LCE is
\begin{equation}
X_r(x_1,...,x_r;n_1,...,n_r) = \sum_{\cG \in \cI^{(r,0)}}
    \sum_{x_{r+1},...,x_v}
    {V_0^{(e)}(n_1,...,n_r;\cG) \, V_0^{(i)}(\cG) \, 
     L_0(x_1,...,x_v; \cG) \over S(\cG)},
\label{X-0}
\end{equation}
where $\cI^{(r,0)}$ is the set of all $r$-rooted connected graphs.

$X$ is invariant under simultaneous permutation of coordinates and
valences:
\[
X_r(x_1,...,x_r;n_1,...,n_r) =
X_r(x_{\pi(1)},...,x_{\pi(r)};n_{\pi(1)},...,n_{\pi(r)}),
\]
but not over independent permutation of coordinates and valences.
Furthermore, $X$ is invariant over the lattice symmetry group, e.g.\
it is translation-invariant:
\[
X_r(x_1,...,x_r;n_1,...,n_r) =
X_r(x_1+x,...,x_r+x;n_1,...,n_r).
\]
$X_1(x;n)$ is independent of $x$, and it will be denoted by $X_1(n)$;
it will also be denoted by $\mu(n)$ in its role of renormalized
vertex.  $X_2(x_1,x_2;n_1,n_2)$ only depends on the difference
$x_2-x_1$, and it will be denoted by $X_2(x_2-x_1;n_1,n_2)$.  Since,
by invariance under space inversion, $X_2(x;n_1,n_2) =
X_2(-x;n_1,n_2)$, we also have $X_2(x;n_1,n_2) = X_2(x;n_2,n_1)$.

The sum over the location of internal vertices of the $\theta$
functions is by definition the (free) lattice embedding number of
$\cG$ with fixed roots:
\begin{equation}
\sum_{x_{r+1},...,x_v} \prod_{l=1}^e \theta(x_{i(l)}-x_{f(l)}) =
E(x_1,...,x_r; \cG).
\end{equation}
Therefore
\begin{equation}
X_r(x_1,...,x_r;n_1,...,n_r) =
    \sum_{\cG \in \cI^{(r,0)}}
    {V_0^{(e)}(n_1,...,n_r;\cG) \, V_0^{(i)}(\cG) \, 
     K^{e(\cG)} \, E(x_1,...,x_q; \cG) \over S(\cG)}.
\label{X-0a}
\end{equation}

Finally, the $q$-point functions are computed as:
\begin{eqnarray}
G_q(x_1,...,x_q) = 
\sum_{\rm partitions}
X_r(x_{i_{11}},...,x_{i_{r1}}; u_1,...,u_r)
\prod_{l=1}^r \delta_{u_l}(x_{i_{l1}},...,x_{i_{lu_l}}),
\label{G-X}
\end{eqnarray}
where the $q$-point delta function is
\[
\delta_1(x_1) = 1,\quad
\delta_2(x_1,x_2) = \delta(x_1-x_2), \quad ..., \quad
\delta_q(x_1,...,x_q) = \prod_{l=2}^q \delta(x_1-x_l)
\]
and $\{\{i_{11},...,i_{1u_1}\},...,\{i_{r1},...,i_{ru_r}\}\}$ is a
generic partition of $\{1,...,q\}$ into $r$ sets of size $u_1$, ...,
$u_r$.  We will call a root bearing a factor of $\delta_u$ a 
{\em $u$-th order} root.

The two-point function is simply
\[
G_2(x) = X_2(x;1,1) + \delta(x)\,X_1(2).
\]

\section{Vertex-renormalized expansion}
\label{1-renorm}

The graph $\cG \backslash i$ is obtained by {\em deleting\/} from
$\cG$ the vertex $i$, i.e.\ by removing $i$ and all edges incident with
$i$.  A vertex $i$ of a rooted graph $\cG$ is called an {\em
articulation point\/} if there exist vertices of $\cG \backslash i$
not connected to a root.  A rooted graph is called 1-{\em
irreducible\/} if it does not contain any articulation point.

\begin{figure}[tb]
\centerline{\psfig{file=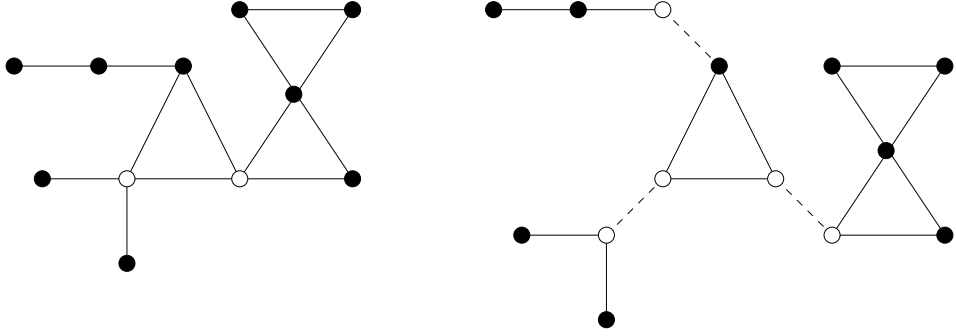}}
\caption{
Example of decomposition of a graph into its 1-skeleton and
1-decorations.  Open dots are roots, filled dots are internal
vertices.}
\label{fig-1ren}
\end{figure}

Any $r$-rooted ($r>1$) connected graph $\cG$ can be decomposed in a
unique way into a 1-irreducible $r$-rooted 1-{\em skeleton\/} $\cS$
and a 1-rooted 1-{\em decoration} for each vertex; $\cG$ is
reconstructed by decorating each vertex, identifying the root of its
decoration with the vertex; an example is presented in Fig.\ 
\ref{fig-1ren}.

Since the only 1-irreducible 1-rooted graph is the single-vertex
graph, we use a different definition for 1-rooted graphs.  A 1-rooted
graph is called a 1-{\em skeleton\/} if it has no articulation points
{\em except the root}.  Any 1-rooted connected graph $\cG$ can be
decomposed in a unique way into a 1-rooted 1-{\em skeleton\/} $\cS$
and a 1-rooted 1-{\em decoration} for each vertex {\em except the
root}, which is left undecorated.

A 1-rooted connected graph is called a 1-{\em insertion\/} if
$\cG \backslash 1$ is connected.

The LCE can be reorganized by summing together all contributions from
graphs having the same 1-skeleton, incorporating 1-decorations into
{\em renormalized vertices\/} $\mu(n) = X_1(n)$ (named {\em
semi-invariants\/} and denoted by $M_n$ in Ref.\ \cite{Wortis};
$\mu(n)=(n-1)!!\,m_n$ in the notations of Ref.\ 
\cite{Luscher-Weisz}).  The unrenormalized LCE of $\mu(n)$ is given by 
Eq.\ (\ref{X-0}).

The $r$-point function can be computed restricting the sum in Eq.\ 
(\ref{X-0}) or (\ref{X-0a}) to the (much smaller) set $\cI^{(r,1)}$ of
1-irreducible $r$-rooted graphs:
\begin{eqnarray}
X_r(x_1,...,x_r;n_1,...,n_r) &=& \sum_{\cG \in \cI^{(r,1)}}
    \sum_{x_{r+1},...,x_v}
    {V_1^{(e)}(n_1,...,n_r;\cG) \, V_1^{(i)}(\cG) \, 
     L_0(x_1,...,x_v; \cG) \over S(\cG)} \nonumber \\
       &=& \sum_{\cG \in \cI^{(r,1)}}
    {V_1^{(e)}(n_1,...,n_r;\cG) \, V_1^{(i)}(\cG) \,
     K^{e(\cG)} \, E(x_1,...,x_r; \cG) \over S(\cG)},
\label{G-1}
\end{eqnarray}
where the internal and external renormalized vertex factors are
\begin{equation}
V_1^{(e)}(n_1,...,n_r;\cG) = \prod_{i=1}^r \mu(n(i)+n_i)
\end{equation}
and
\begin{equation}
V_1^{(i)}(\cG) = \prod_{i=r+1}^v \mu(n(i)).
\label{V-1i}
\end{equation}

Eq.\ (\ref{X-0}) requires a sum over all connected graphs, and
therefore it is impractical for the computation of $\mu(n)$ at large
orders of the LCE.  We introduce the {\em renormalized moments\/}
$q(n)$ (named {\em self-fields\/} and denoted by $G_n$ in Ref.\ 
\cite{Wortis}), defined by
\begin{equation}
q(n) = \sum_{\cG \in \cI^{(1,0)}_{n,\rm in}} \sum_{x_2,...,x_v}
    {V_0^{(i)}(\cG) \, L_0(x_1,...,x_v; \cG) \over S(\cG)},
\end{equation}
where $\cI^{(1,0)}_{n,\rm in}$ is the set of 1-insertions with root of
valence $n$.  The following equations hold:
\begin{equation}
q(n) =  \sum_{\cG \in \cI^{(1,1)}_{n,\rm in}}
    \sum_{x_2,...,x_v}
    {V_1^{(i)}(\cG) \, L_0(x_1,...,x_v; \cG) \over S(\cG)},
\label{q-1}
\end{equation}
where $\cI^{(1,1)}_{n,\rm in} = \cI^{(1,0)}_{n,\rm in} \cap
\cI^{(1,1)}$, and $\cI^{(1,1)}$ is the set of 1-rooted 1-skeletons;
\begin{equation}
\mu(n) = 
  \mu_0(n) + \sum_{s=1}^\infty {1\over s!} \sum_{l_1=1}^\infty ...
    \sum_{l_s=1}^\infty q(l_1)\,...\,q(l_s) \, \mu_0(n+l_1+...+l_s).
\label{mu-1}
\end{equation}
Since $q(2n-1)=0$, and odd values of $n$ and $l_i$ do not contribute
to Eq.\ (\ref{mu-1}).  $q$ and $\mu$ can now be computed recursively
in parallel order by order in $K$, since, once Eq.\ (\ref{q-1}) is
expanded in powers of $K$ and truncated, the coefficient of the
highest power of $K$ of $q$ in the l.h.s.\ depends only on
coefficients of lower powers of $K$ of $\mu$ in the r.h.s.

\section{Edge-renormalized expansion}
\label{2-renorm}

A pair of distinct vertices $i$ and $j$ of a rooted graph $\cG$ is
called an {\em articulation pair\/} if there exist vertices of 
$\cG \backslash i \backslash j$ not connected to a root, or if $i$ and
$j$ are joined by more than one edge.  A rooted graph is called 
2-{\em irreducible\/} if it does not contain any articulation pair.

\begin{figure}[tb]
\centerline{\psfig{file=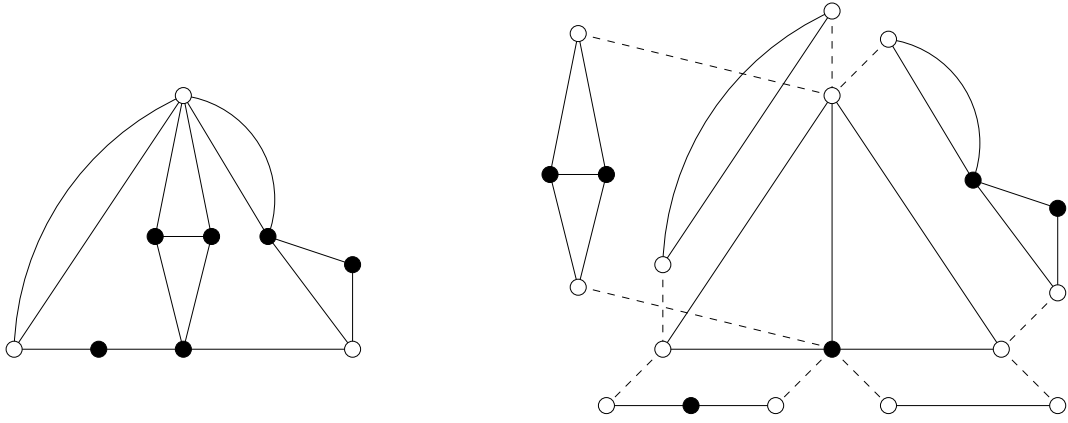}}
\caption{
Example of decomposition of a graph into its 2-skeleton and
2-decorations.  Open dots are roots, filled dots are internal
vertices.}
\label{fig-2ren}
\end{figure}

Any 1-irreducible $r$-rooted ($r>2$) graph $\cG$ can be decomposed in
a unique way into a 2-irreducible $r$-rooted 2-{\em skeleton\/} $\cS$
and a 1-irreducible 2-rooted 2-{\em decoration} for each edge
(oriented in a canonical way, e.g.\ by choosing $i(l)<f(l)$); $\cG$ is
reconstructed by replacing each edge with its decoration, identifying
the first and second decoration root with the initial and final vertex
of the edge respectively.  An example is shown in
Fig.\ \ref{fig-2ren}.

We use a different definition for 2-rooted graphs, since the only
2-irreducible 2-rooted graph is the {\em bond\/} graph (no internal
vertices and only one edge); the roots of all other 1-irreducible
2-rooted graphs are an articulation pair.  We call a 1-irreducible
2-rooted graph a 2-{\em skeleton\/} if it does not contain any
articulation pair {\em except the pair consisting of the two roots}.
Any 1-irreducible 2-rooted graph can be decomposed in a unique way
into a 2-rooted 2-{\em skeleton\/} $\cS$ and a 1-irreducible 2-rooted
2-{\em decoration} for each edge {\em except edges connecting the
roots}, which are left undecorated.

The LCE can be reorganized by summing together all contributions from
graphs having the same 2-skeleton, incorporating all 2-decorations
into {\em renormalized edges}.

We start by decomposing Eq.\ (\ref{G-1}) for $r>1$ into
\begin{eqnarray}
W_r(x_1,...,x_r;n_1,...,n_r) &=&
    \sum_{\cG \in \cI^{(r,1)}_{n_1,...,n_r}}
    \sum_{x_{r+1},...,x_v}
    {V_1^{(i)}(\cG) \, L_0(x_1,...,x_v; \cG) \over S(\cG)},
\label{G-valence} \\
X_r(x_1,...,x_r;n_1,...,n_r) &=&
    \sum_{s_1,...,s_r} \Biggl[\prod_{i=1}^r \mu(n_i+s_i)\Biggr] 
    W_r(x_1,...,x_r;s_1,...,s_r),
\label{X-W}
\end{eqnarray}
where $\cI^{(r,1)}_{n_1,...,n_r}$ is the set of 1-irreducible
$r$-rooted graphs with roots of valence $n_1$, ..., $n_r$.  $W_r$
enjoys the same symmetry properties of $X_r$. $W_1(x,n)$ is undefined.

$W_r$ can be computed by assigning an initial and a final valence
$i_l$, $f_l$ to each oriented edge of a 2-rooted graph; the valence
$i_l$ is incident with $i(l)$ and $f_l$ is incident with $f(l)$:
\begin{eqnarray}
&&W_r(x_1,...,x_r;n_1,...,n_r) = \sum_{\cG \in \cI^{(r,2)}}
    \sum_{x_{r+1},...,x_v;i_1,...,i_e,f_1,...,f_e} 
\prod_{i=1}^r \delta(n_i-\nu_i(i_1,...,i_e,f_1,...,f_e;\cG)) \,
\nonumber \\
&&\quad\times\; {V_2^{(i)}(i_1,...,i_e,f_1,...,f_e;\cG) \, 
         L_2(x_1,...,x_v; i_1,...,i_e,f_1,...,f_e; \cG) 
         \over S(\cG)},
\label{G2-r}
\end{eqnarray}
where $\cI^{(r,2)}$ is the set of 2-irreducible $r$-rooted graphs,
$\nu_i$ is the sum of all the valences incident with the vertex $i$,
\begin{equation}
V_2^{(i)}(i_1,...,i_e,f_1,...,f_e;\cG) =
 \prod_{i=r+1}^v \mu(\nu_i(i_1,...,i_e,f_1,...,f_e;\cG)),
\end{equation}
and
\begin{equation}
L_2(x_1,...,x_v; i_1,...,i_e,f_1,...,f_e; \cG) = 
\prod_{l=1}^e W_2(x_{i(l)}-x_{f(l)};i_l,f_l).
\label{L-2}
\end{equation}

For $r=2$ Eqs.\ (\ref{G2-r}) and (\ref{L-2}) require a slight
modification: we define $\cI^{(2,2)}$ as the set of 2-rooted
2-skeletons; for edges incident with at least one root, we replace
$W_2(x_1-x_2;i,f)$ with $K \theta(x_1-x_2) \delta_{i1} \delta_{f1}$.

The sum over graphs can be restricted to a subset of $\cI^{(r,2)}$; we
will discuss here the case $r=2$; the next Section will be devoted to
the case $r\ge3$.  Let us start by classifying 1-irreducible 2-rooted
graphs into several classes.

An internal vertex $i$ of a 1-irreducible 2-rooted graph $\cG$ is
called a {\em nodal point\/} if the roots of $\cG \backslash i$ are
not connected.  A 1-irreducible 2-rooted graph is {\em nodal\/} (also
named {\em articulated\/} or {\em separable\/} in the literature) if
it contains one or more nodal points; otherwise it is {\em non-nodal}.

A 1-irreducible 2-rooted graph is {\em simple\/} if $\cG \backslash 1
\backslash 2$ is connected, and 1 is not adjacent to 2.  By
definition, all nodal graphs are simple.  A 1-irreducible 2-rooted
graph is a {\em ladder\/} graph if it is not simple and it is not the
bond graph.

A 1-irreducible 2-rooted graph is {\em elementary\/} if it is both
simple and non-nodal.

We have divided 1-irreducible 2-rooted graphs into four disjoint
classes: bond, nodal, ladder, and elementary graphs.  Let us separate
the contributions to $W_2$ according to the four classes:
\[
W_2(x;n_1,n_2) = W^{\rm bo}_2(x;n_1,n_2) + W^{\rm no}_2(x;n_1,n_2) + 
                 W^{\rm la}_2(x;n_1,n_2) + W^{\rm el}_2(x;n_1,n_2),
\]
i.e.\ bond, nodal, ladder, and elementary contributions respectively.

The bond contribution is trivial.  Nodal contributions can be
factorized into a product of non-nodal contributions:
\begin{eqnarray}
&&W^{\rm no}_2(x;n_1,n_2) = 
\sum_{x_3;i_1,i_2} W^{\rm nn}_2(x_3;n_1,i_1) \,
    \mu(i_1+i_2) \, W^{\rm nn}_2(x-x_3;i_2,n_2) \nonumber \\
&&\;+\;
\sum_{x_3,x_4;i_1,i_2,i_3,i_4} W^{\rm nn}_2(x_3;n_1,i_1) \,
    \mu(i_1+i_2)\nonumber \\
&&\quad\times\; W^{\rm nn}_2(x_4-x_3;i_2,i_3) \,
    \mu(i_3+i_4) \, W^{\rm nn}_2(x-x_4;i_4,n_2) + ... \, ,
\end{eqnarray}
which can be written recursively as
\begin{equation}
W^{\rm no}_2(x;n_1,n_2) = 
\sum_{x_3;i_1,i_2} W^{\rm nn}_2(x_3;n_1,i_1) \,
    \mu(i_1+i_2) \, W_2(x-x_3;i_2,n_2).
\label{Gno-2}
\end{equation}

Likewise, ladder contributions can be factorized into a product of
non-ladder contributions:
\begin{equation}
W^{\rm la}_2(x;n_1,n_2) = 
\sum_{s=2}^\infty {1\over s!} \sum_{i_1,...,i_s,f_1,...,f_s}
\delta\Bigl(n_1-{\textstyle \sum_{t=1}^s} i_t\Bigr) \,
\delta\Bigl(n_2-{\textstyle \sum_{t=1}^s} f_t\Bigr) \,
\prod_{t=1}^s W^{\rm nl}_2(x,i_t,f_t).
\label{Gla-2}
\end{equation}

Elementary contributions can be computed by setting $r=2$ into Eq.\ 
(\ref{G2-r}), and restricting the sum to $\cI^{(2,2)}_{\rm el}$, the
set of {\em elementary\/} 2-rooted 2-skeletons.  The sum can be
further restricted to $\cI^{(2,2)}_{0, \rm el}$, the set of {\em
unordered\/} elementary 2-rooted 2-skeletons, provided that we replace
$S(\cG)$ with $S_E(\cG)$ and we symmetrize the result:
\begin{eqnarray}
&&W^{\rm el}_2(x;n_1,n_2) = {1\over2}
    \sum_{\cG \in \cI^{(2,2)}_{0, \rm el}}
    \sum_{x_3,...,x_v;i_1,...,i_e,f_1,...,f_e} 
\prod_{l=1}^2 \delta(n_i-\nu_i(i_1,...,i_e,f_1,...,f_e;\cG)) \,
\nonumber \\
&&\quad\times\; {V_2^{(i)}(i_1,...,i_e,f_1,...,f_e;\cG) \, 
         L_2(x_1,...,x_v; i_1,...,i_e,f_1,...,f_e; \cG) 
         \over S_E(\cG)} + n_1 \leftrightarrow n_2.
\label{G2el-2}
\end{eqnarray}

The last ingredient we need for a fully edge-renormalized expansion
are the renormalized vertices $\mu(n)$; they can be computed by
combining Eq.\ (\ref{mu-1}) with
\begin{equation}
q(n_1{+}n_2) = {n_1!\,n_2!\over(n_1+n_2)!} 
(W^{\rm no}_2(0;n_1,n_2) + W^{\rm el}_2(0;n_1,n_2)),
\label{q-2}
\end{equation}
reflecting the fact that the contributions to $q(n_1{+}n_2)$ in Eq.\ 
(\ref{q-1}) can be obtained from the contributions to $W_2(0;n_1,n_2)$
in Eq.\ (\ref{G-valence}) by {\em identifying\/} the roots of the
2-rooted graph, i.e.\ by suppressing the second root and reattaching
all the edges incident with it to the first root, provided that the
roots are not adjacent.

Expanding in powers of $K$ Eqs.\ (\ref{Gno-2}), (\ref{Gla-2}),
(\ref{G2el-2}), (\ref{q-2}), and (\ref{mu-1}), we can compute 
$W^{\rm el}_2(x;n_1,n_2)$, $W^{\rm no}_2(x;n_1,n_2)$, 
$W^{\rm la}_2(x;n_1,n_2)$, $q(n)$, and $\mu(n)$ in parallel order by
order in $K$. The only step which involves a summation over graphs is
Eq.\ (\ref{G2el-2}), where the sum only runs over $\cI^{(2,2)}_{0, \rm
el}$, a relatively small set.

\section{Three-point and higher functions}
\label{q-point}

A vertex (internal or external) $i$ of a 1-irreducible $r$-rooted
graph $\cG$ is called a {\em nodal point\/} if $\cG \backslash i$ is
not connected.  By definition of 1-irreducibility, each connected
component of $\cG \backslash i$ must contain at least one root.  A
1-irreducible $r$-rooted graph $\cG$ is {\em nodal\/} if it contains
one or more nodal points; otherwise it is {\em non-nodal}.

In the rest of this section, we will assume that every graph is
2-irreducible.

A nodal point $j$ of a 2-irreducible $r$-rooted graph $\cG$ is called
a {\em tree-insertion point\/} if at least one of the connected
components of $\cG \backslash j$ is a tree graph.  The order $t$ of
$j$ is the number of roots of $\cG$ contained in all tree graph
components of $\cG \backslash j$, plus 1 if $j$ itself is a root.  A
2-irreducible $r$-rooted graph is called {\em compact\/} if it
contains no tree-insertion points.

\begin{figure}[tb]
\centerline{\psfig{file=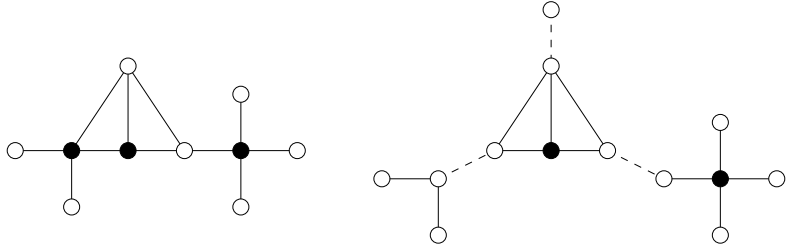}}
\caption{
Decomposition of a 2-irreducible graph into a compact kernel and
2-irreducible tree graphs.
}
\label{fig-Wred}
\end{figure}

Let us consider a 2-irreducible $r$-rooted graph $\cG$.  We generate a
compact 2-irreducible graph $\cG'$, called the {\em compact kernel\/}
of $\cG$, by removing all the tree graphs attached to every
tree-insertion point, and promoting all internal tree-insertion points
to root.  $\cG$ is obtained by attaching a 2-irreducible tree graph to
each root of $\cG'$.  If $\cG'$ is not a tree graph (and therefore it
has at least 3 roots), the decomposition is unique; otherwise, $\cG$
itself is a tree graph.  An example is shown in Fig.\ \ref{fig-Wred}.

By summing all contribution of 2-irreducible graphs with the same
compact kernel, we can write $W_q$ as
\begin{eqnarray}
&&W_q(x_1,...,x_q;n_1,...,n_q) =
W^{\rm tr}_q(x_1,...,x_q;n_1,...,n_q) \nonumber \\
&&\ +\,\sum_{\textstyle{{\rm partitions}\atop r>2}}
\sum_{y_1,...,y_r,i_1,...,i_r} W^{\rm co}_r(y_1,...,y_r;i_1,...,i_r)
\prod_{l=1}^r  Y_{u_l+1}(y_l,x_{i_{l1}},...,x_{i_{lu_l}};
                         i_l,n_{i_{l1}},...,n_{i_{lu_l}}),
\nonumber \\
\label{Wq}
\end{eqnarray}
where $W^{\rm tr}_q(x_1,...,x_q;n_1,...,n_q)$ is the
tree graph contribution to $W_q$, 
$W^{\rm co}_r(y_1,...,y_r;i_1,...,i_r)$ is the compact graph
contribution to $W_r$, and
\begin{eqnarray}
Y_{t+1}(y,x_1,...,x_t;i,n_1,...,n_t) &=& 
\sum_j \mu(i+j) \, W^{\rm tr}_{t+1}(y,x_1,...,x_t;j,n_1,...,n_t) 
\nonumber \\
&+& \sum_j \delta(y-x_1)\,\delta(n_1-i-j)\,
W^{\rm tr}_t(x_1,...,x_t;j,n_2,...,n_t), \nonumber \\
Y_2(y,x;i,n) &=& \sum_j \mu(i+j) \, W^{\rm tr}_2(y,x;j,n) 
+ \delta(y-x)\,\delta(n-i)
\end{eqnarray}
(the first and second term correspond to an internal and external
$t$-tree-insertion point respectively); notice that 
$W^{\rm tr}_2=W_2$.  Eq.\ (\ref{Wq}) can be combined with Eq.\ 
(\ref{G-X}) to give
\begin{eqnarray}
&&G_q(x_1,...,x_q) = 
G^{\rm tr}_q(x_1,...,x_q) \nonumber \\
&&\ +\,\sum_{\textstyle{{\rm partitions}\atop r>2}}
\sum_{y_1,...,y_r,i_1,...,i_r} W^{\rm co}_r(y_1,...,y_r;i_1,...,i_r)
\prod_{l=1}^r 
Z_{u_l+1}(y_l,x_{i_{l1}},...,x_{i_{lu_l}};i_l),
\label{G-Z}
\end{eqnarray}
where
\begin{eqnarray}
Z_{t+1}(y,x_1,...,x_t;i) &=&
\sum_{\textstyle{{\rm partitions}\atop{\rm of}\ \{1,...,t\}}}
\sum_{s_1,...,s_r} Y_{r+1}(y,x_{i_{11}},...,x_{i_{r1}}; i,s_1,...,s_r)
\nonumber \\
&&\quad\times\ 
\prod_{l=1}^r \mu(s_l+u_l)\,\delta_{u_l}(x_{i_{l1}},...,x_{i_{lu_l}}).
\end{eqnarray}
$Z_{t+1}(y,x_1,...,x_t;i)$ is symmetric under permutations of
$x_1,...,x_t$ and lattice symmetries, e.g.\ simultaneous translation
of $y$ and $x_1,...,x_t$.

These formulae can be written graphically, according to the
rules presented in Table \ref{tab-rules}.  A sum over all dummy
coordinates $y$ and all dummy valences $i,f$ and a sum over
all inequivalent permutation of external coordinates $x$ or
coordinate-valence pairs $x,n$ are understood.  Notice that, despite
the graphical notation, all pairs of roots of $W^{\rm nn}_q$ and
$W^{\rm co}_q$ are equivalent.

\begin{table}[tbp]
\caption{
Graphical rules.  $\nu$ is the sum of all the valences incident with
the vertex.  The argument $z_i$ of $W^{\rm nn}_q$ is the coordinate
$x$ or $y$ associated with the symbol placed at the vertex $i$ of the
polygon; the argument $y_i$ of $W^{\rm co}_q$ is the variable $y$ of
the function $Z_{q+1}$ placed at the vertex $i$.
}
\label{tab-rules}
\begin{center}
\begin{tabular}{|c|c|c|c|}
\hline
symbol & comment & variables & factor \\
\hline
\psfig{file=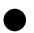} & internal vertex & $y$ & $\mu(\nu)$ \\
\psfig{file=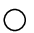} & unlabelled root & $x$, $n$ & $\delta(n-\nu)$ \\
\lower1ex\psfig{file=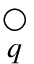} & labelled root & 
$x_1,...,x_q$ & $\mu(q+\nu)\,\delta_q(x_1,...,x_q)$ \\
\lower1ex\psfig{file=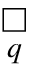} & first root of $Z$ & 
$x_1,...,x_q$, $n$ & $\mu(q+\nu+n-1)\,\delta_q(x_1,...,x_q)$ \\
\lower1ex\psfig{file=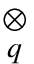} & root of $G$ & $y,x_1,...,x_q$ &
$Z_{q+1}(y,x_1,...,x_q;\nu)$ \\
\lower1ex\vbox{\psfig{file=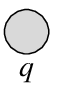}\vskip-1ex\null} & & $x_1,...,x_q$ &
$G^{\rm tr}_q(x_1,...,x_q)$ \\ 
\psfig{file=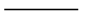} & & $i,f$ & $W_2(x_{i(l)}-x_{f(l)};i,f)$ \\
\multicolumn{2}{|c|}{$q$-sided polygon} & &
$W^{\rm nn}_q(z_1,...,z_q;\nu_1,...,\nu_q)$ \\
\multicolumn{2}{|c|}{$q$-sided polygon with a letter ``c''} & &
\vphantom{\lower 2pt \hbox{$W_Q$}}%
$W^{\rm co}_q(y_1,...,y_q;\nu_1,...,\nu_q)$ \\
\hline
\end{tabular}
\end{center}
\end{table}

\begin{figure}[tb]
\centerline{\psfig{file=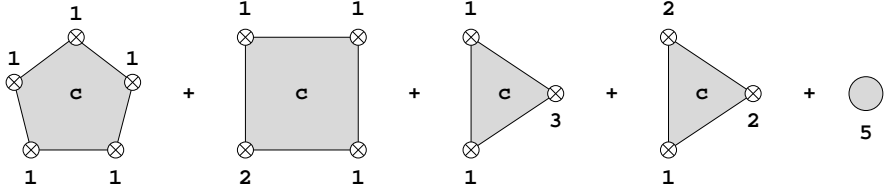}}
\caption{
Graphical representation of Eq.\ (\protect\ref{G-Z}) for $q=5$.
}
\label{fig-G5}
\end{figure}

Eq.\ (\ref{G-Z}) can be expressed by writing all polygons with a
letter ``c'' having 3 to $q$ vertices, placing a crossed dot with a
positive integer label at each vertex in all inequivalent ways, the
sum of the labels being $q$, and adding the tree contribution.  The
case $q=5$ is shown in Fig.\ \ref{fig-G5}.

\begin{figure}[tb]
\centerline{\psfig{file=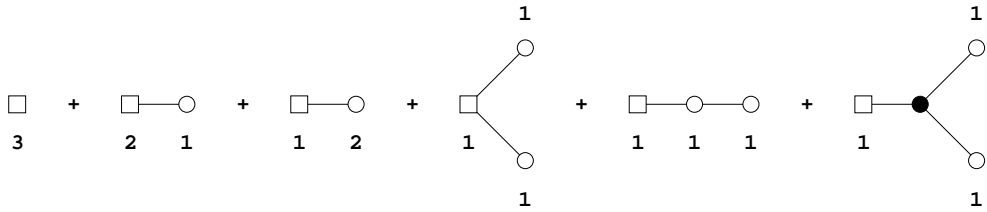}}
\caption{
Contributions to $Z_3$.
}
\label{fig-Z3}
\end{figure}

$G_q$ and $Z_q$ can be computed by adding the contributions of all
2-irreducible $r$-rooted tree graphs with roots labelled by positive
integers with sum $q$; for $Z_q$, the first root must be drawn as a
square.  The case $Z_3$ is shown in Fig.\ \ref{fig-Z3}.

\begin{figure}[tb]
\centerline{\psfig{file=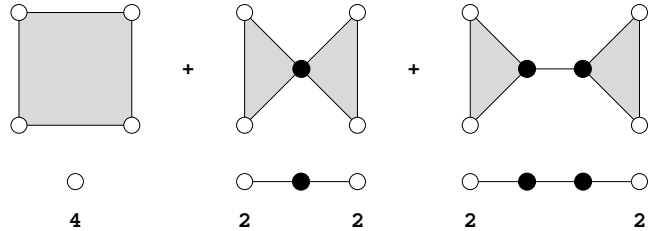}}
\caption{
Compact contributions to $W^{\rm co}_4$ and corresponding nodal
skeletons.
}
\label{fig-W4co}
\end{figure}

The next step is to write an expression of $W^{\rm co}_q$ in terms of
$W^{\rm nn}_r$.  For $q=3$ we have simply $W^{\rm co}_3 = W^{\rm
nn}_3$.  The case $q=4$ is shown in Fig.\ \ref{fig-W4co}.
For larger values of $q$, the number of non-nodal contributions to
$W^{\rm co}_q$ grows rapidly, and a systematic approach is needed.

Let us define for a (connected or non-connected) graph $\cG$ and a
vertex $i$ the graph $\cG/i$: let $\cG^*$ be the connected component
of $\cG$ containing $i$; if $\cG^* \backslash i$ is connected, set
$\cG/i = \cG$; otherwise, for each connected component $\cG^*_k$ of
$\cG^* \backslash i$ generate the graph ${\overline\cG}_k$ by adding a
new vertex, internal or external like $i$, and joining it to all the
vertices adjacent to $i$ in $\cG^*$, by the same number of edges;
replace $\cG^*$ with the connected components ${\overline\cG}_k$.  The
edges and vertices of $\cG/i$ are in one-to-one correspondence with
the edges and vertices of $i$, except for the new vertices which all
correspond to $i$ (a nodal point of $\cG$).  Notice that
$\cG/i/j=\cG/j/i$.

Let $\cG$ be a compact 2-irreducible $r$-rooted graph ($r\ge3$) with
$t$ nodal points $i_1,...,i_t$.  Observe that all the connected
components $\cG_l$ of $\overline\cG \equiv \cG/i_1/.../i_t$ are
non-nodal.  Generate a new graph $\cT$, the {\em nodal skeleton\/} of
$\cG$, in the following way: for each $\cG_l$ with $v\ge3$, containing
$n$ roots corresponding to non-nodal roots of $\cG$, write a root $l$
of $\cT$ with a label $n\ge0$; for each node $i_k$ of $\cG$ write an
unlabelled vertex $i_k$ of $\cT$, internal or external like $i_k$;
join $i_k$ to all the labelled roots $l$ such that $\cG_l$ contain a
vertex corresponding to $i_k$, and with all unlabelled vertices which
are adjacent to $i_k$ in $\cG$.

A nodal skeleton is a connected 1-irreducible tree graph, but it is
not in general 2-irreducible (it may contain 2-valent internal
vertices).  A nodal skeleton enjoys the following properties: each
2-valent internal vertex is adjacent to a labelled root; labelled roots 
are never adjacent; unlabelled roots are at least 2-valent;
$m$-valent roots with label $n$ satisfy $n+m\ge3$.  Every nodal
skeleton can be generated by adding labels to some of the roots and by 
splicing 2-valent internal vertices into a 2-irreducible tree graph.

Every 1-irreducible tree graph, with some roots carrying a
non-negative integer label, satisfying the above properties, is the
nodal skeleton of a set of $q$-rooted 2-irreducible compact graphs,
with $q$ equal to the sum of the labels plus the number of unlabelled
roots.  The contribution of this set to $W^{\rm co}_q$ can be computed 
by replacing each $m$-valent root of $\cT$ carrying a label $n$ with
an $(n+m)$-sided polygon whose vertices are the $m$ vertices adjacent
to the root and $n$ new (unlabelled) roots, and applying the rules
of Table \ref{tab-rules}.

\begin{figure}[tb]
\centerline{\psfig{file=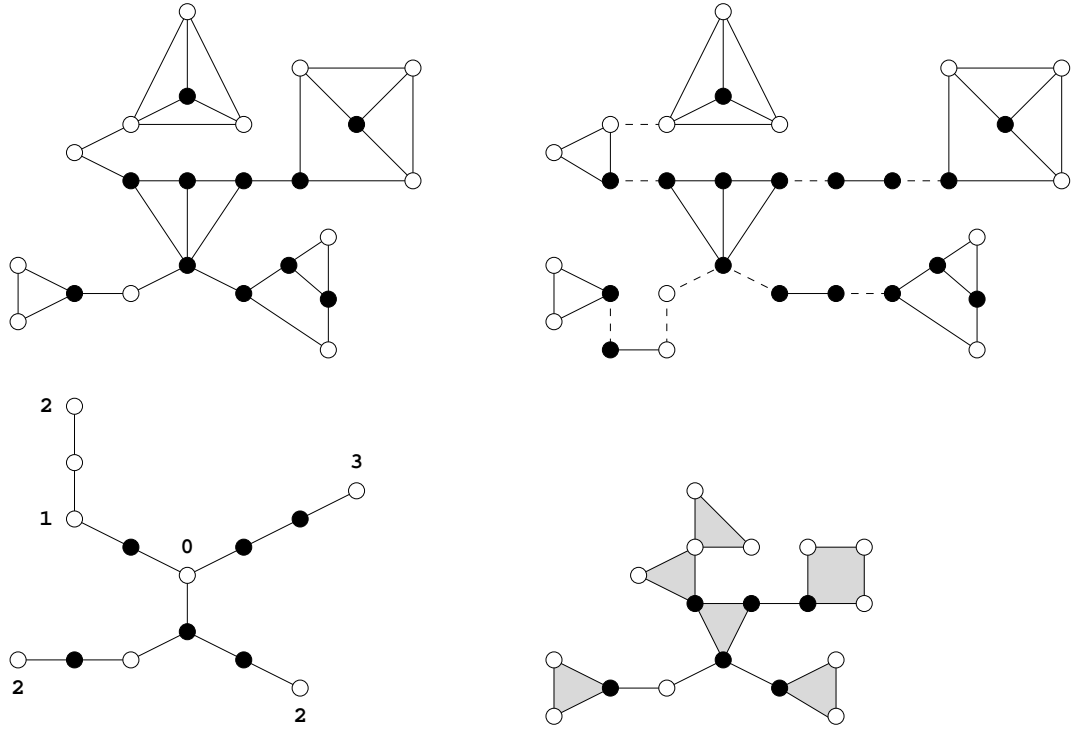}}
\caption{A compact graph, its non-nodal components, its nodal
skeleton, and the contribution to $W^{\rm co}$ of the set of graphs
sharing the nodal skeleton.
}
\label{fig-nodskel}
\end{figure}

\begin{figure}[tb]
\centerline{\psfig{file=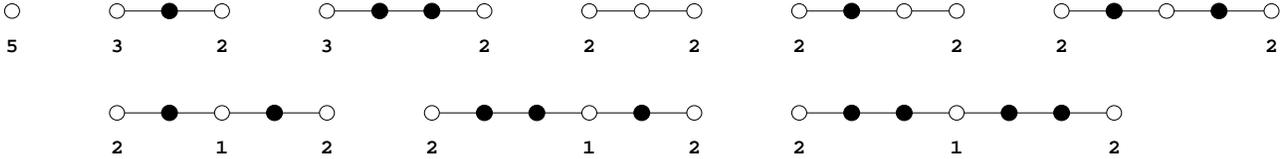,width=170truemm}}
\caption{
Nodal skeletons contributing to $W^{\rm co}_5$.
}
\label{fig-W5co}
\end{figure}

An example of the construction of the nodal skeleton and its
evaluation is presented in Fig.\ \ref{fig-nodskel}.  The set of all
nodal skeletons contributing to $W^{\rm co}_5$ is shown in Fig.\
\ref{fig-W5co}; see also Fig.\ \ref{fig-W4co}.

We could carry further the reduction of the set of graphs to be summed
over, e.g.\ by identifying ladder graphs along the lines of Sect.\ 
\ref{2-renorm}.  This is rather complicated for arbitrary $r$, and
goes beyond the scope of the present work.  Moreover, the
zero-momentum projection described below is not applicable to the
ladder graph reduction.  Therefore we compute $W^{\rm nn}_r$ by
restricting the sum of Eq.\ (\ref{G2-r}) to $\cI^{(r,2)}_{\rm nn}$,
the set of {\em non-nodal\/} 2-irreducible $r$-rooted graphs.

The above considerations can be simplified considerably if we are only 
interested in moments of the $q$-point functions, e.g.\ 
\begin{eqnarray}
\chi_q &\equiv& \sum_{x_2,...,x_q} G_q(x_1,...,x_q), \nonumber \\
M_2^{(q)} &\equiv& \sum_{x_2,...,x_q} (x_1-x_2)^2 G_q(x_1,...,x_q).
\label{m2q}
\end{eqnarray}
Let us introduce the moments of $Z_q$, $W_q^{\rm co}$ and 
$W_q^{\rm nn}$:
\begin{eqnarray}
\zeta_q(n) &\equiv& \sum_{x_2,...,x_q} Z_q(x_1,...,x_q;n), \nonumber \\
\omega^{\rm co}_q(n_1,...,n_q) &\equiv& 
\sum_{x_2,...,x_q} W^{\rm co}_q(x_1,...,x_q;n_1,...,n_q),
\end{eqnarray}
the corresponding definition for $\chi^{\rm tr}_q$ and
$\omega^{\rm nn}_q(n_1,...,n_q)$, all independent second moments, etc.

Eq.\ (\ref{G-Z}) can be projected over zero momentum to give
\begin{equation}
\chi_q = \chi^{\rm tr}_q + \sum_{r=3}^q 
    \sum_{u_1 \le...\le u_r \atop u_1 + ... + u_r = q} 
    {q! \over \prod_{l=1}^r u_l!}
    \sum_{i_1,...,i_r} 
    \omega^{\rm co}_r(i_1,...,i_r)
    \prod_{l=1}^r \zeta_{u_l+1}(i_l).
\end{equation}

The computation of $\zeta_q$, $\chi^{\rm tr}_q$, and 
$\omega^{\rm co}_q$ is also easy; the graphical rules can be
immediately projected over zero momentum, suppressing all coordinates
$x$ and $y$ and removing all space delta functions; the only
nontrivial part is the counting of the number of inequivalent
permutations of roots.

$\zeta_q(n)$ can be computed by summing over all inequivalent
2-irreducible 1-ordered $r$-rooted tree graphs, with the roots
labelled by positive integers $u_1,...,u_r$ with sum $q$.  The number
of inequivalent permutations of the roots 2, ..., $r$ is
\[
{(q-1)! \over S (u_1-1)! \prod_{l=2}^r u_l!},
\]
where $S$ is the symmetry factor of the labelled graph.

The computation of $\chi^{\rm tr}_q(n)$ is very similar, but we sum over
unordered graphs, and the number of inequivalent permutations of the
roots is 
\[
{q! \over S \prod_l u_l!}.
\]

$\omega^{\rm co}_q(n_1,...,n_q)$ can be computed by summing over all
inequivalent unordered $r$-rooted nodal skeletons, with $p$ roots
labelled by non-negative integers $u_1,...,u_p$ with 
$r-p+u_1+...+u_p = q$, with a weight $1/S$, where $S$ is the symmetry
factor of the labelled graph, with unlabelled roots assigned an
arbitrary distinct label (e.g. $-1$).

Finally, we can compute $\omega^{\rm nn}_q(n_1,...,n_q)$ by summing
over unordered non-nodal 2-irreducible $q$-rooted graphs, provided
that we use the modified symmetry factor and we symmetrize the result
under permutation of the valences.

The computation of the second moment of the above quantities proceeds
along the same lines.  The factor $(x_i-x_j)^2$ in Eq.\ (\ref{m2q}) is
dealt with in the following way: the two roots are connected by a
chain of terms with a space structure of the form
\[
\sum_{x_j,y_1,...,y_n} (x_i-x_j)^2 
f_1(x_i-y_1)\,f_2(y_1-y_2)\, ... \,f_{n+1}(y_n-x_j)
\]
(we have dropped the dependency on coordinates lying outside the 
branch connecting $i$ with $j$).  Let us write
\[
(x_i-x_j)^2 = (x_i-y_1)^2 + (y_1-y_2)^2 + ... + (y_n-x_j)^2 + 
\hbox{cross terms}.
\]
The cross terms do not contribute to the sum, and the result is
\[
f_1^{(2)} f_2^{(0)} ... f_{n+1}^{(0)} +
f_1^{(0)} f_2^{(2)} ... f_{n+1}^{(0)} + ... +
f_1^{(0)} f_2^{(0)} ... f_{n+1}^{(2)},
\]
where
\[
f_i^{(0)} = \sum_y f_i(y), \qquad f_i^{(2)} = \sum_y y^2 f_i(y).
\]
Therefore we can compute the second moment by taking each contribution
to the zero-momentum quantity, promoting one of the zero-momentum
factors along the branch connecting the roots $i$ and $j$ to second
moment, and summing over all possible choices.

By dealing with moments, we avoid the need of storing all the values
of $Z$, $W^{\rm co}$, and $W^{\rm nn}$, which can rapidly exhaust all
available memory.  The extension to higher moments is straightforward
but cumbersome.

\section{Programming details}
\label{Programming}

We wrote a set of computer programs to implement the automatic
evaluation of the edge-renormalized LCE on the simple cubic lattice
(sc) and on the body-centered cubic lattice (bcc) ($3D$); the same
programs evaluate the LCE on two different representations of the
square lattice ($2D$) and on the $1D$ lattice.

The computation of $q$-point functions is performed for a generic
potential, keeping $\mu_0(2n)$ symbolic; each term of the series is a
polynomial in $\mu_0(2n)$ with rational coefficients.  We also
implemented the same computation for a specific potential; this
requires much less memory and is somewhat faster (up to 30\%), but not
enough to give up the flexibility of a generic potential.

To speed up search and insertion into ordered sets of data, graph sets
and polynomials in $\mu_0(2n)$ are implemented as AVL trees
(height-balanced binary trees) (cfr.\ e.g.\ Ref.\ \cite{Knuth},
Chapt.\ 6.2.3), using the ubiqx library.  Rational numbers and
(potentially) large integers are handled by the GNU multiprecision
(gmp) library.

Given the complexity of the procedure, it is crucial to perform a
number of checks in order to flush out all algorithm and program
errors.  In $1D$ our series are compared with exact results for the
spin-1/2 \cite{CPRV-4point} and the spin-1 \cite{Rossi-priv} Ising
model; this is already a very stringent check, especially of the graph
sets (cfr.\ Ref.\ \cite{Nickel-Rehr}).  In $2D$, our results are
compared with the series for $\chi$ and $M_2$ for spin-1/2 published
in Ref.\ \cite{Nickel-Cargese}.  In $3D$, our results are compared
with the lower-order series already available, for $\chi$ and $M_2$
for specific potentials in Refs.\ 
\cite{Nickel-Cargese,Nickel-Rehr,Butera-Comi}, and for $\chi$, $M_2$,
and $\chi_q$ for a generic potential in Ref.\ \cite{CPRV-Ising-es}.
$q(n)$ can be computed from different combinations of $n_1$ and $n_2$
in Eq.\ (\ref{q-2}); their agreement is non-trivial.  Finally, for the
spin-1/2 Ising model on any lattice, the series for $G_q$, rewritten
in terms of $v=\tanh K$, must have integer coefficients.

\subsection{Graph generation}

A program generates the table of all unordered elementary 2-rooted
2-skeletons contributing to the desired order; the algorithm follows
Ref.\ \cite{Nickel-Rehr}.  Starting from the graph drawn in Fig.\
\ref{fig-element}, we apply recursively the following modified Heap
rules \cite{Heap}:

\begin{figure}[tb]
\centerline{\psfig{file=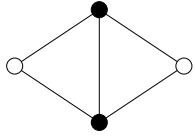}}
\caption{
The simplest elementary 2-rooted 2-skeleton.
}
\label{fig-element}
\end{figure}

(a) join any two distinct vertices by a new edge, 
{\em provided the two vertices are not already adjacent\/};

(b) insert a new internal vertex on any edge and join it to any
vertex, {\em excluding the edge extrema}, by a new edge;

(c) insert two new internal vertices on any two {\em distinct} edges,
and join them by a new edge;

(d) {\em do not join the roots by a new edge}.

The modifications to the original Heap rules (a), (b), and (c), marked
in italics, prevent the generation of 2-reducible graphs.  Rule (d)
prevents the generation of ladder graphs.

The reduction of graphs to canonical form is performed using a
generalization of the algorithm of Ref.\ \cite{Nickel-Rehr}.  Graphs
are stored in a compact form similar to the one of Ref.\ 
\cite{Nickel-Rehr}.

To reduce the proliferation of graphs at higher orders, it is
extremely important to know the order (``strict bound'') $o_s$ at
which a given graph will enter in the expansion (it is not trivially
$e$, since we require even valence of all internal vertices, and,
being interested in bipartite lattices, even length of all loops).

We must also keep in mind that some graphs don't contribute at the
desired order, but graphs generated from them might contribute.  We
define the ``Heap bound'' $o_H(\cG)$ as the minimum of $o_s$ on the
set of graphs including $\cG$ and all graphs generated from it.  We
also define the two bounds for $(\cG,\eta)$, i.e.\ the minimal order
when the vertex $i$ is forced to be embedded in a lattice site of
parity $\eta(i)$.

We apply the modified Heap rules to the graph $\cG$ in the following
way: assign a parity label $\eta(i)$ to each vertex, in all the ways
compatible with the Heap bound; apply the modified Heap rules
assigning all possible parity labels to the new vertices; discard
immediately the generated graph-parity pairs not satisfying the Heap
bound; discard vertex parity information and store the generated
graphs not isomorph to previously generated graphs.  Finally, save
into a file only the graphs satisfying the strict bound.

The generation of the elementary 2-rooted 2-skeletons contributing to
the 25th order required 41 hours of computation on one CPU of a Compaq
ES-40, and ca.\ 300 Mbytes of RAM.  The number of inequivalent
elementary 2-skeletons for each order of $o_s$ is reported in Table
\ref{tab-2skel}.

\begin{table}[tbp]
\caption{
Number of inequivalent unordered elementary 2-rooted 2-skeletons
($r=2$), or unordered non-nodal 2-irreducible $r$-rooted graphs
($r>2$), for each order of the strict bound $o_s$.  The number of
graphs satisfying the Heap bound is typically 20\% to 30\% higher.
No graph in any of these sets has a bound lower than 4.
}
\label{tab-2skel}
\begin{center}
\begin{tabular}{|c|r|r|r|r|r|r|r|}
\hline
$o_s$ & \multicolumn{1}{|c|}{$r=2$}   
      & \multicolumn{1}{|c|}{$r=3$}   
      & \multicolumn{1}{|c|}{$r=4$}
      & \multicolumn{1}{|c|}{$r=5$}   
      & \multicolumn{1}{|c|}{$r=6$}
      & \multicolumn{1}{|c|}{$r=7$}
      & \multicolumn{1}{|c|}{$r=8$} \\
\hline
\04 &       0 &       1 &       1 &       0 &       0 &       0 &       0 \\
\05 &       0 &       0 &       0 &       0 &       0 &       0 &       0 \\
\06 &       0 &       0 &       1 &       2 &       1 &       0 &       0 \\
\07 &       0 &       1 &       1 &       1 &       1 &       0 &       0 \\
\08 &       1 &       3 &       5 &       4 &       4 &       2 &       1 \\
\09 &       0 &       2 &       4 &       6 &       6 &       5 &       2 \\
 10 &       3 &       7 &      19 &      26 &      27 &      22 &      12 \\
 11 &       0 &       9 &      23 &      47 &      63 &      48 &      33 \\
 12 &      13 &      46 &     111 &     175 &     229 &     228 &     159 \\
 13 &       6 &      54 &     168 &     378 &     603 &     661 &     575 \\
 14 &      59 &     263 &     737 &    1436 &    2224 &    2691 &    2465 \\
 15 &      29 &     367 &    1364 &    3473 &    6404 &    8694 &    9216 \\
 16 &     367 &    1855 &    5824 &   13190 &   23766 &   34106 &   38239 \\
 17 &     197 &    2898 &   12088 &   34726 &   72900 &  116210 &  146284 \\
 18 &    2589 &   14937 &   51801 &  133739 &  275031 & & \\
 19 &    1547 &   25332 &  118225 &  375859 &  884317 & & \\
 20 &   21682 &  135325 &  514319 & & & & \\
 21 &   13933 &  245306 & 1251818 & & & & \\
 22 &  199865 & & & & & & \\
 23 &  139610 & & & & & & \\
 24 & 2026682 & & & & & & \\
 25 & 1516576 & & & & & & \\
\hline
\end{tabular}
\end{center}
\end{table}

A similar program generates all unordered non-nodal 2-irreducible
$r$-rooted graphs for $r\ge3$.  The table is initialized by applying
rules (a) and (b) recursively, starting from each 2-irreducible
$r$-rooted tree graph, until the result is non-nodal.  Rules (a), (b),
and (c) are then applied recursively.  The number of inequivalent
non-nodal 2-irreducible $r$-rooted graphs for each order of $o_s$ is
reported in Table \ref{tab-2skel}.

We remark that this graph tables can be used for the LCE of any spin
model with $\phi\to-\phi$ symmetry on a bipartite lattice in any
dimension.

The generation of all the required families of tree graphs is
straightforward.

\subsection{Computation of the $q$-point functions}

A separate program reads the table of elementary 2-rooted 2-skeletons
and computes all components of $W_2$.  

The evaluation of $\mu$ and of bond, nodal, and ladder contribution to
$W_2$ is a straightforward application of the formulae of Sect.\ 
\ref{2-renorm}.

The evaluation of elementary contribution dominates the computation
time, and must be optimized as much as possible.  Assume that all
lower-order contributions to $W_2$ have been computed.  For each
unordered elementary 2-skeleton $\cG$ with $o_s$ not larger than the
desired order, all inequivalent assignations $\cG,(n,l)$ of edge
valence parity $n$ and length parity $l$ compatible with the desired
order, with even length of all loops, and with even valence of all
internal vertices, are generated.  We have implemented two different
algorithms for the computation of the contribution of $\cG,(n,l)$ to
the two-point function.

The first algorithm is essentially the one used by Nickel and Rehr in
Ref.\ \cite{Nickel-Rehr}: all inequivalent 1-irreducible 2-rooted
graphs with a 2-skeleton compatible with $\cG,(n,l)$ are generated,
and their contributions are computed according to Eq.\ (\ref{G-1}).
In the second algorithm, the contribution is computed according to
Eq.\ (\ref{G2el-2}), and 1-irreducible graphs are not needed.  The
first algorithm is more efficient for 2-skeletons with large $o_s$,
while the second algorithm is more efficient for small $o_s$; for each
value of $o_s$ we select the algorithm which is (presumably) more
efficient.  On the sc lattice, the speed-up obtained over the use of
either algorithm for all skeletons grows with the order, and is about
a factor of 4 at order 23.  On the bcc lattice, the first algorithm is
very efficient, since the embedding number factorizes into a product
of 1-dimensional embedding numbers \cite{Nickel-Rehr}; we still use
the second algorithm for the simplest 2-skeletons ($o_s\le10$), since
the computation of the corresponding 1-irreducible 2-rooted graphs
contributing to orders higher than 21 is extremely time- and
memory-consuming.

Keeping in RAM all components of $W_2$ for a generic potential would
be problematic.  Most of these components are needed only to compute
nodal and ladder contributions, and can be kept on disk; keeping in
RAM just the components needed to compute elementary contribution is
manageable.

The computation of the 25th-order LCE for the two-point function on
the bcc lattice required ca.\ 400 hours of computation, and ca.\ 700
Mbytes of RAM.

A similar program reads the table of non-nodal 2-irreducible
$r$-rooted graphs and the components of $W_2$, and computes
$\omega^{\rm nn}_q$.  The computation of $\chi_q$ is then
straightforward.  We computed $\chi_4$, $\chi_6$, and $\chi_8$ to 21st,
19th, and 17th order respectively on the bcc lattice.

The computation of the same quantities on the sc lattice is much
slower (but does not requires more RAM); so far, we obtained $W_2$ to
23th order, with an effort not much smaller than the 25th order on the
bcc lattice.  The computation of $W_2$ and $\chi_q$ to the same orders
as on the bcc lattice is in progress, but it will require a
non-trivial amount of time.

\section{Selected results}
\label{Results}

All high-temperature series computed in the present work are available
for the most general potential, in the form of polynomials in the bare
vertices $\mu_0(2n)$.  The general results are extremely lengthy, and
are only useful for further computer processing.

We present here a selection of high-temperature series for the
spin-1/2 Ising model, i.e.\ for
\[
f(\phi) = \half\Bigl(\delta(\phi+1) + \delta(\phi-1)\Bigr).
\]
For sake of compactness, all the series are written in terms of
$v=\tanh K$.  Series for other specific potentials are available upon
request from the author.

Although we computed all components of $G_2(x,y)$, we report here only
$\chi\equiv\chi_2$ and $M_2\equiv M_2^{(2)}$ (cfr.\ Eq.\ (\ref{m2q})).
For the $q$-point functions, we only computed $\chi_q$.

On the bcc lattice, we obtained

\vbox{
\begin{eqnarray}
\chi &=& 
  1 + 8\,v + 56\,{v^2} + 392\,{v^3} + 2648\,{v^4} + 17864\,{v^5} + 
  118760\,{v^6} + 789032\,{v^7} 
\nonumber \\ &&\; +\, 
  5201048\,{v^8} + 34268104\,{v^9} + 
  224679864\,{v^{10}} + 1472595144\,{v^{11}} + 9619740648\,{v^{12}} 
\nonumber \\ &&\; +\,
  62823141192\,{v^{13}} + 409297617672\,{v^{14}} + 
  2665987056200\,{v^{15}} + 
  17333875251192\,{v^{16}}
\nonumber \\ &&\; +\,
  112680746646856\,{v^{17}} + 
  731466943653464\,{v^{18}} + 4747546469665832\,{v^{19}} 
\nonumber \\ &&\; +\,
  30779106675700312\,{v^{20}} + 199518218638233896\,{v^{21}} + 
  1292141318087690824\,{v^{22}}
 \nonumber \\ &&\; +\,
  8367300424426139624\,{v^{23}} + 
  54141252229349325768\,{v^{24}} 
 \nonumber \\ &&\; +\,
  350288350314921653160\,{v^{25}} +
   {\rm O}(v^{26});
\end{eqnarray}
}
\vbox{
\begin{eqnarray}
M_2 &=& 
  8\,v + 128\,{v^2} + 1416\,{v^3} + 13568\,{v^4} + 119240\,{v^5} + 
  992768\,{v^6} + 7948840\,{v^7} 
\nonumber \\ &&\; +\, 
  61865216\,{v^8} + 470875848\,{v^9} + 
  3521954816\,{v^{10}} + 25965652936\,{v^{11}}
\nonumber \\ &&\; +\, 
  189180221184\,{v^{12}} + 1364489291848\,{v^{13}} +
  9757802417152\,{v^{14}}
\nonumber \\ &&\; +\, 
  69262083278152\,{v^{15}} + 488463065172736\,{v^{16}} +
  3425131086090312\,{v^{17}}
\nonumber \\ &&\; +\, 
  23896020585393152\,{v^{18}} + 
  165958239005454632\,{v^{19}} +  1147904794262960384\,{v^{20}}
 \nonumber \\ &&\; +\, 
  7910579661767454248\,{v^{21}} + 54332551216709931904\,{v^{22}}
\nonumber \\ &&\; +\, 
  372033905161237212392\,{v^{23}} + 2540342425838560175616\,{v^{24}}
\nonumber \\ &&\; +\, 
  17301457207110720278440\,{v^{25}} +
   {\rm O}(v^{26});
\end{eqnarray}
}
\vbox{
\begin{eqnarray}
-\chi_4 &=& 
  2 + 64\,v + 1168\,{v^2} + 16576\,{v^3} + 201232\,{v^4} +
  2204608\,{v^5} + 22411504\,{v^6}
\nonumber \\ &&\; +\,
  215447872\,{v^7} + 1981980688\,{v^8} + 
  17602809920\,{v^9} + 151865668752\,{v^{10}}
\nonumber \\ &&\; +\,
  1278888344256\,{v^{11}} + 10550227820400\,{v^{12}} +
  85510907958720\,{v^{13}}
 \nonumber \\ &&\; +\,
  682500568307184\,{v^{14}} + 5374496030148928\,{v^{15}} + 
  41821018545214608\,{v^{16}}
\nonumber \\ &&\; +\,
  321992795063663936\,{v^{17}} + 2455641803116052752\,{v^{18}} +
  18567879503614668736\,{v^{19}}
\nonumber \\ &&\; +\,
  139310655514229882000\,{v^{20}} + 1037854026688655887552\,{v^{21}} +
   {\rm O}(v^{22});
\end{eqnarray}
}
\vbox{
\begin{eqnarray}
\chi_6 &=& 
16 + 1088\,v + 36416\,{v^2} + 853952\,{v^3} + 15974528\,{v^4} + 
  255491264\,{v^5} + 3638767040\,{v^6}
\nonumber \\ &&\; +\,
  47395195712\,{v^7} + 
  574950589568\,{v^8} + 6581949043264\,{v^9} + 71803170318144\,{v^{10}}
\nonumber \\ &&\; +\,
  752047497945024\,{v^{11}} + 7606707093034368\,{v^{12}} + 
  74649010982738112\,{v^{13}}
\nonumber \\ &&\; +\,
  713458387977120192\,{v^{14}} + 
  6661638582474716480\,{v^{15}} + 60923519621981242752\,{v^{16}} 
\nonumber \\ &&\; +\,
  546923327751320201536\,{v^{17}} + 4828463182433394315584\,{v^{18}}
\nonumber \\ &&\; +\,
  41987611565592990702272\,{v^{19}} + 
   {\rm O}(v^{20});
\end{eqnarray}
}
\vbox{
\begin{eqnarray}
-\chi_8 &=& 
272 + 31744\,v + 1673728\,{v^2} + 58110976\,{v^3} + 1538207872\,{v^4} + 
  33584739328\,{v^5}
\nonumber \\ &&\; +\,
  634387677184\,{v^6} + 10699575811072\,{v^7} + 
  164723097021568\,{v^8} 
\nonumber \\ &&\; +\,
  2352360935459840\,{v^9} + 
  31540880634427392\,{v^{10}} + 400802365468148736\,{v^{11}}
\nonumber \\ &&\; +\,
  4862781935250449280\,{v^{12}} + 56665753776838026240\,{v^{13}}
\nonumber \\ &&\; +\,
  637305912177206767104\,{v^{14}} + 6945658883867865975808\,{v^{15}}
\nonumber \\ &&\; +\,
  73600395257678784586368\,{v^{16}} + 760476823195422275111936\,{v^{17}} +
    {\rm O}(v^{18}).
\nonumber \\
\end{eqnarray}
}

On the sc lattice, we obtained

\vbox{
\begin{eqnarray}
\chi &=& 
  1 + 6\,v + 30\,{v^2} + 150\,{v^3} + 726\,{v^4} + 3510\,{v^5} + 
  16710\,{v^6} + 79494\,{v^7} + 375174\,{v^8}
\nonumber \\ &&\; +\,
  1769686\,{v^9} + 8306862\,{v^{10}} + 
  38975286\,{v^{11}} + 182265822\,{v^{12}} + 852063558\,{v^{13}}
\nonumber \\ &&\; +\,
  3973784886\,{v^{14}} + 18527532310\,{v^{15}} +
  86228667894\,{v^{16}} + 401225368086\,{v^{17}}
\nonumber \\ &&\; +\,
  1864308847838\,{v^{18}} + 8660961643254\,{v^{19}} +
  40190947325670\,{v^{20}}
\nonumber \\ &&\; +\,
  186475398518726\,{v^{21}} + 864404776466406\,{v^{22}} +
  4006394107568934\,{v^{23}}
\nonumber \\ &&\; +\,
  {\rm O}(v^{24});
\end{eqnarray}
}
\vbox{
\begin{eqnarray}
M_2 &=& 
  6\,v + 72\,{v^2} + 582\,{v^3} + 4032\,{v^4} + 25542\,{v^5} +
  153000\,{v^6} + 880422\,{v^7} + 4920576\,{v^8}
\nonumber \\ &&\; +\,
  26879670\,{v^9} + 144230088\,{v^{10}} + 
  762587910\,{v^{11}} + 3983525952\,{v^{12}} + 20595680694\,{v^{13}}
\nonumber \\ &&\; +\,
  105558845736\,{v^{14}} + 536926539990\,{v^{15}} +
  2713148048256\,{v^{16}} + 13630071574614\,{v^{17}}
\nonumber \\ &&\; +\,
  68121779384520\,{v^{18}} + 338895833104998\,{v^{19}} +
  1678998083744448\,{v^{20}}
\nonumber \\ &&\; +\,
  8287136476787862\,{v^{21}} + 40764741656730408\,{v^{22}} + 
  199901334823355526\,{v^{23}}
\nonumber \\ &&\; +\,
  {\rm O}(v^{24}).
\end{eqnarray}
}

\subsection*{Acknowledgments}

We would like to thank Michele Caselle and Ettore Vicari for many
useful discussions.  Special thanks are due to Paolo Rossi, who worked
out exact results for the $1D$ spin-1 Ising model, and to Andrea
Pelissetto, for his critical reading of the manuscript.


\end{document}